\documentclass[aps,prl,superscriptaddress,twocolumn,showpacs]{revtex4}
\usepackage{amsmath}
\usepackage{graphicx}
\usepackage{amssymb}
\usepackage{bbm, color, soul}
\usepackage{graphicx}
\usepackage{adjustbox}

\usepackage{tikz}
\usepackage{pgfplots}
\pgfplotsset{compat=newest}

\usepackage{color}
\usepackage{multirow}

\usepackage{tikz}

\usepackage{amsthm}%%% it must be load
\usepackage{tikz-cd}
\usepackage[all]{xy}
\usepackage{amsfonts}
\usepackage{mathrsfs}%%% to use \mathscr
\usepackage{hyperref}%%% for hyperlinks
\usepackage{MnSymbol} % provides the ``end of def'' symbol
\theoremstyle{plain}

\theoremstyle{definition}
\newtheorem{defn/}[equation]{Definition}

\theoremstyle{remark}
\newtheorem{rmk/}[equation]{Remark}
\newtheorem*{rmk*}{Remark}

\newtheorem*{claim*}{Claim}

\begin{document}

\title[]{Chaos-controlled switching between entanglement and coherence}

\author{Kyu-Won \surname{Park}}
\email{parkkw7777@gmail.com}
\affiliation{Department of Mathematics and Research Institute for Basic Sciences, Kyung Hee University, Seoul, 02447, Korea}

\author{Soojoon \surname{Lee}}
\email{level@khu.ac.kr}
\affiliation{Department of Mathematics and Research Institute for Basic Sciences, Kyung Hee University, Seoul, 02447, Korea}
\affiliation{School of Computational Sciences, Korea Institute for Advanced Study, Seoul 02455, Korea}

\author{Kabgyun \surname{Jeong}}
\email{kgjeong6@snu.ac.kr}
\affiliation{Research Institute of Mathematics, Seoul National University, Seoul 08826, Korea}
\affiliation{School of Computational Sciences, Korea Institute for Advanced Study, Seoul 02455, Korea}

\date{\today}

% =========================================================
% Abstract
% =========================================================

\begin{abstract}
Controlling entanglement and coherence is central to quantum information, yet the two resources often exhibit antagonistic trends and are difficult to optimize within a single platform. Here we show that chaos enables \emph{switchable} eigenstate resources: avoided crossings in soft- versus strong-chaos windows selectively realize an entanglement-peak mode or a coherence-peak mode within the same system. Crucially, this chaos-controlled inversion is not tied to a particular notion of subsystems, appearing both in single-wave settings and in genuine many-body settings. From the quantum-chaos perspective, conventional diagnostics based on avoided-crossing phenomenology and eigenmode delocalization are insufficient; eigenfunction entanglement and basis coherence provide the missing discriminants.
Using two wave-chaotic billiards and a tilted-field Ising chain, we track the information-theoretic response of eigenstates across localized hybridization windows. Even when avoided-crossing phenomenology and delocalization are comparable, the entanglement and coherence responses invert between soft- and strong-chaos regimes. In the Ising chain, a single microscopic knob, the global field tilt, toggles between the two operating modes and reveals a trade-off in which off-diagonal correlations grow as diagonal populations dip. Our diagnostics require only reduced states (or their spectra) and are compatible with mode imaging in wave-chaos resonators and randomized measurements in programmable spin simulators.
\end{abstract}

\maketitle

% =========================================================
% Introduction
% =========================================================
\section{Introduction}
Quantum chaos is commonly diagnosed through spectral and eigenfunction signatures: dense avoided-crossing structures under smooth parameter sweeps \cite{Landau1932,Zener1932,Stockmann1999}, which provide a parametric manifestation of level repulsion, and statistical universality of spectral fluctuations described by random-matrix theory (in contrast to integrable level clustering) \cite{Wigner1951,Bohigas1984,Mehta2004,BerryTabor1977}, and basis delocalization of eigenmodes in space \cite{Berry1977,Gutzwiller1990,Haake2010}, including scarring \cite{Heller1984,Sridhar1991}. These diagnostics capture how classical mixing reshapes wave patterns (for example in dispersing and stadium billiards) \cite{Sinai1970,Bunimovich1979}, but they do not directly address a question that is central in quantum information: how controllable hybridization redistributes \emph{operational resources}, notably entanglement and coherence.

Entanglement and coherence are both indispensable, yet they serve different tasks and are often in tension. Bipartite entanglement enables nonlocal protocols and distributed processing \cite{NielsenChuang2000,Horodecki2009,PlenioVirmani2007,Amico2008}, whereas basis coherence underlies phase-sensitive tasks and robustness to dephasing, with operational meanings in coherence distillation and discrimination \cite{Baumgratz2014,StreltsovRMP2017,WinterYang2016,Napoli2016}. In many settings, increasing mixing raises entanglement while washing out basis coherence, making joint optimization challenging.

A key subtlety is that entanglement depends on how subsystems are defined: it is specified only after choosing a tensor-product structure (TPS), which is in practice induced by accessible observables and controls \cite{Zanardi2004TPS}. This places wave-chaotic billiards and spin chains on the same footing: the former realizes intra-system (inter-degree-of-freedom) entanglement of single-wave eigenmodes, while the latter realizes genuine inter-particle entanglement \cite{vanEnk2005SPE,Azzini2020SPE}. Our universality claim is thus not TPS-independence of the entanglement \emph{value}, but TPS-robustness of the \emph{resource flow} across avoided crossings, namely the chaos-controlled inversion between entanglement and basis coherence.

In quantum billiards, prior work has largely emphasized that stronger chaoticity correlates with increased nonseparability (``classical entanglement'') \cite{Joseph2015,Joseph2016}. More generally, entanglement generation has been explored as a quantum signature of chaos in paradigmatic models \cite{MillerSarkar1999,Wang2004,Chaudhury2009}, and in isolated many-body systems the onset of chaos underlies thermalization scenarios and eigenstate-based statistical mechanics \cite{Deutsch1991,Rigol2008}. Here we show a qualitatively different and sharper statement: \emph{within parameter windows where hybridization and delocalization look similarly chaotic}, the \emph{information-theoretic} response can bifurcate. Specifically, for two wave-chaotic billiards that both display clear mode exchange and configuration-space Shannon-entropy peaks at avoided crossings, the coordinate entanglement can respond with opposite signs. Extending beyond single-particle wave chaos, we then demonstrate in a many-body Ising chain that a single control parameter can toggle between entanglement-enhancing and coherence-enhancing hybridization windows, effectively providing a minimal ``resource switch'' within one Hamiltonian.

The paper is organized as follows. We first introduce the billiard setting and establish the common delocalization signature across avoided crossings. We then resolve the Schmidt spectra and show the peak-versus-dip inversion of entanglement, supported by a compact perturbative interpretation. Next, we complement entanglement by a basis-resolved coherence monotone and a diagonal/off-diagonal purity-channel decomposition, which together reveal whether changes arise from mixedness (entanglement) or from genuine reshaping of off-diagonal structure. Finally, we map the same phenomenology onto a tilted-field Ising chain, where a single knob controls soft-chaos and strong-chaos windows and exposes channel competition in a strongly mixed background, with direct implications for quantum resource control.

% =========================================================
% Quantum-billiard setting
% =========================================================

\section{Quantum-billiard setting}
\label{sec:qbill_setup}

A (closed) billiard provides a minimal arena where geometric constraints alone generate complex dynamics. Classically, a point particle moves freely inside a bounded domain and undergoes specular reflection at the hard wall. This is encoded by
\begin{equation}
\mathcal{H}(q,p)=\frac{p^{2}}{2m}+V(q),\qquad
V(q)=
\begin{cases}
0, & q\in\Omega,\\
\infty, & q\in\partial\Omega,
\end{cases}
\label{eq:class_billiard_H}
\end{equation}
which generates an area-preserving flow on phase space and becomes mixing for the deformed geometries of interest \cite{Sinai1970,Bunimovich1979}.

Quantum mechanically, stationary modes are obtained from the Dirichlet problem
\begin{equation}
\begin{cases}
(\nabla^{2}+n^{2}k^{2})\psi(\mathbf r)=0, & \mathbf r\in\Omega,\\
\psi(\mathbf r)=0, & \mathbf r\in\partial\Omega,
\end{cases}
\label{eq:helmholtz_dirichlet}
\end{equation}
where \(k\) is the vacuum wavenumber and \(nk\) is the internal wavenumber for refractive index \(n\).
Throughout we adopt TM polarization and take \(n=3.3\), so that \(\psi\) represents the out-of-plane field component.
The resulting eigenpairs \(\{k,\psi(\mathbf r)\}\) define the spectral branches used below.

\paragraph{Deformation parameters.}
We use two distinct geometries and reserve distinct deformation parameters:
\begin{equation}
\eta=\kappa \quad \text{(quadrupole)},\qquad
\eta=\varepsilon \quad \text{(oval)}.
\label{eq:eta_kappa_epsilon}
\end{equation}
Whenever a formula applies to both billiards, we write it in terms of the dummy control parameter \(\eta\).

\paragraph{Hilbert-space viewpoint.}
Let \(\Omega\subset\mathbb{R}^{2}\) and consider \(\mathcal{H}=L^{2}(\Omega)\) with inner product
\begin{equation}
\langle \phi,\psi\rangle=\int_{\Omega}\phi^{*}(\mathbf r)\psi(\mathbf r)\,d^{2}\mathbf r ,
\label{eq:H_L2}
\end{equation}
so each eigenmode is a normalized vector in \(\mathcal{H}\).
Numerically, we represent \(\psi(x,y)\) on a rectangular grid, which induces a finite-dimensional tensor product structure used to define reduced states and Schmidt spectra in Sec.~\ref{sec:entanglement_recap}.

% =========================================================
% Classical entropy / delocalization (Fig. 1)
% =========================================================
\section{Avoided crossings and configuration-space delocalization}
\label{sec:phase-space}

To generate controlled avoided crossings \cite{Landau1932,Zener1932} we consider two deformed hard-wall geometries with different symmetry and mixing strength.
The soft-chaos reference is the quadrupole billiard with boundary
\begin{equation}
\rho(\theta)=1+\kappa\cos(2\theta),
\label{eq:quad_boundary}
\end{equation}
where \(\kappa=0\) recovers the circle and both mirror symmetries \(x\to -x\), \(y\to -y\) are preserved.

As a strong-chaos counterpart we study an oval billiard obtained by deforming an ellipse in the \(x\) direction,
\begin{equation}
\frac{x^{2}}{a^{2}}+\bigl(1+\varepsilon x\bigr)\frac{y^{2}}{b^{2}}=1,
\label{eq:oval_boundary}
\end{equation}
with \(a=1.0\) and \(b=1.03\). The deformation \(\varepsilon\) breaks integrability and reduces the discrete symmetry to a single mirror axis (the \(x\)-axis).

\begin{figure*}
\centering
\includegraphics[width=15.5cm]{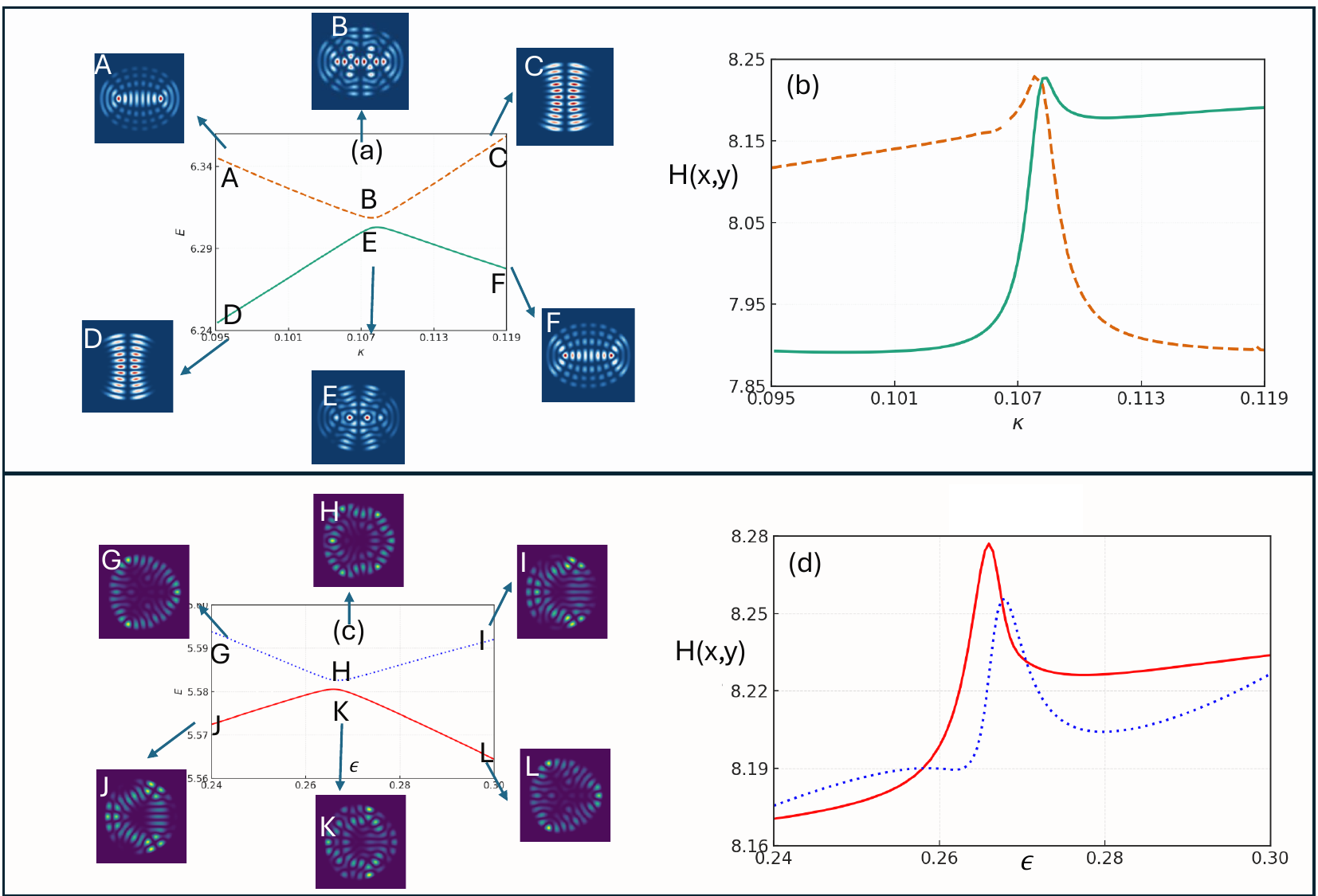}
\caption{\textbf{Avoided crossings and configuration-space Shannon entropy in quadrupole and oval billiards.}
(a,b) Quadrupole: two branches form an avoided crossing as \(\kappa\) is varied; representative intensities show mode exchange, while the configuration-space Shannon entropy peaks in the hybridization window.
(c,d) Oval: the same analysis versus \(\varepsilon\), again showing mode exchange and a Shannon-entropy peak despite stronger background mixing.}
\label{Figure-1}
\end{figure*}

Figure~\ref{Figure-1} establishes a common delocalization baseline. For each eigenmode we normalize the intensity
\begin{equation}
p(x,y)=\frac{|\psi(x,y)|^{2}}{\int_{\Omega}|\psi(x,y)|^{2}\,dx\,dy},
\label{eq:p_xy_norm}
\end{equation}
and compute the configuration-space Shannon entropy \cite{Shannon1948,CoverThomas2006}
\begin{equation}
S_{\mathrm{Sh}}^{xy}(\eta)=-\int_{\Omega} p(x,y)\,\log p(x,y)\,dx\,dy .
\label{eq:SSh_xy}
\end{equation}
In both geometries, the avoided crossing produces a localized enhancement of \(S_{\mathrm{Sh}}^{xy}\), consistent with rapid reshuffling of basis weights during mode exchange.
In what follows we show that this shared delocalization signature does \emph{not} uniquely determine the entanglement and coherence response.

% =========================================================
% Quantum entropy (Fig. 2 / F22)
% =========================================================
\section{Avoided crossings and quantum entropy}

\subsection{Tensor-product structure and coordinate entanglement}
\label{sec:entanglement_recap}
Avoided crossings provide a localized window where two neighboring eigenmodes hybridize, enabling a controlled comparison of quantum correlations along a smooth deformation. 
A key point is that entanglement is defined only after specifying a tensor-product structure (TPS), i.e., what we regard as subsystems~\cite{
Zanardi2004TPS, vanEnk2005SPE, Azzini2020SPE} . In the present work this lets us treat two complementary notions on the same footing: intra-system (inter-degree-of-freedom) entanglement for a single-wave eigenmode in a billiard, versus inter-particle entanglement in many-body spin chains. 
For billiards, we quantify \emph{coordinate (spatial-mode) entanglement} with respect to the bipartition between the \(x\) and \(y\) degrees of freedom.

On the full plane one has \(L^{2}(\mathbb{R}^{2})\simeq L^{2}(\mathbb{R}_x)\otimes L^{2}(\mathbb{R}_y)\).
For a billiard domain \(\Omega\) the exact factorization does not hold geometrically, but our numerical representation on a rectangular grid induces a finite-dimensional factorization,
\begin{equation}
\mathcal{H}_N \simeq \mathbb{C}^{N_x}\otimes \mathbb{C}^{N_y}\equiv \mathcal{H}_x\otimes \mathcal{H}_y,
\label{eq:H_discrete_tensor}
\end{equation}
with
\begin{equation}
|\psi(\eta)\rangle=\sum_{i=1}^{N_x}\sum_{j=1}^{N_y}\psi_{ij}(\eta)\,|x_i\rangle\otimes|y_j\rangle,
\label{eq:psi_tensor}
\end{equation}
normalized by \(\sum_{i,j}|\psi_{ij}|^{2}\Delta x\,\Delta y=1\).
The reduced density matrices are
\begin{equation}
\rho_x(\eta)=\mathrm{Tr}_y\!\left(|\psi(\eta)\rangle\langle\psi(\eta)|\right)\;,
\rho_y(\eta)=\mathrm{Tr}_x\!\left(|\psi(\eta)\rangle\langle\psi(\eta)|\right)
\label{eq:reduced_rhos}
\end{equation}

and the von Neumann entropy
\begin{equation}
S_{\mathrm{vN}}(\eta)=-\mathrm{Tr}\bigl(\rho_x(\eta)\log \rho_x(\eta)\bigr)
\label{eq:SvN_billiard}
\end{equation}
quantifies nonseparability of \(\psi(x,y)\) across the coordinate cut.

\subsection{Schmidt-spectrum reshuffling and opposite entanglement response}
\label{sec:svd_pert}

Let \(C(\eta)\) denote the coefficient matrix built from the discretized eigenmode:
\begin{equation}
C(\eta)=\frac{\psi(\eta)}{\bigl(\sum_{x,y}|\psi(x,y;\eta)|^2\bigr)^{1/2}} .
\label{eq:C_def_eta}
\end{equation}
Then
\begin{equation}
\rho_x(\eta)=C(\eta)C(\eta)^\dagger,\qquad 
\rho_y(\eta)=C(\eta)^\dagger C(\eta),
\label{eq:rho_from_C_eta}
\end{equation}
and the Schmidt coefficients are the squared singular values of \(C(\eta)\).
We compute the leading Schmidt spectrum (rank 5) and the corresponding entropy \(S_{\mathrm{vN}}(\eta)\); in our datasets the leading five weights typically capture more than \(0.99\) of the total Schmidt weight, justifying this truncation for tracking reshuffling trends.

\begin{figure*}[t]
\centering
\includegraphics[width=15.5cm]{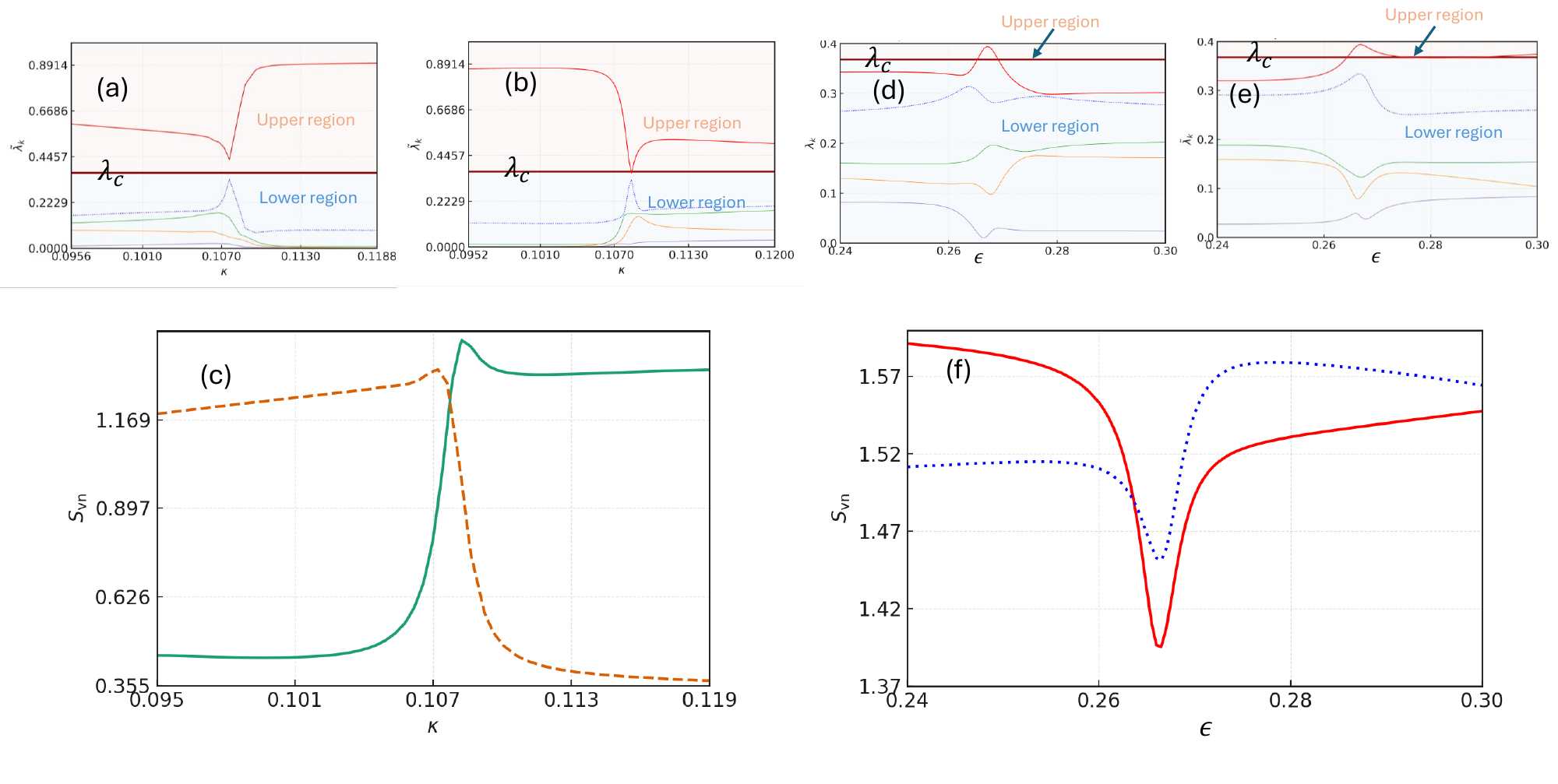}
\caption{\textbf{Schmidt-spectrum reshuffling and entanglement inversion across avoided crossings.}
(a--c) Quadrupole: the leading Schmidt weights flatten across the avoided crossing, producing an entropy peak.
(d--f) Oval: the leading Schmidt weights concentrate across the avoided crossing, producing an entropy dip.
The reference threshold \(\lambda_c=1/e\) highlights whether hybridization promotes subdominant weights (equalization) or sharpens the leading sector (concentration). Across the shown window, the leading five weights typically sum to \(>0.99\), so rank-5 already captures the relevant reshuffling.
}
\label{Figure-2}
\end{figure*}

Figure~\ref{Figure-2} shows the central inversion.
In the quadrupole billiard, hybridization transfers weight from the dominant Schmidt component into subdominant ones, transiently flattening the spectrum and producing a pronounced peak in \(S_{\mathrm{vN}}\).
In the oval billiard, the reshuffling is reversed: weight concentrates into the leading sector and \(S_{\mathrm{vN}}\) develops a clear dip, even though Fig.~\ref{Figure-1} shows a similar delocalization peak in configuration space.
Thus, within avoided-crossing windows that exhibit comparable mode exchange and basis reshuffling, entanglement can respond with opposite signs, implying an information-theoretic structure not captured by delocalization alone.
\begin{figure*}
\centering
\includegraphics[width=16.5cm]{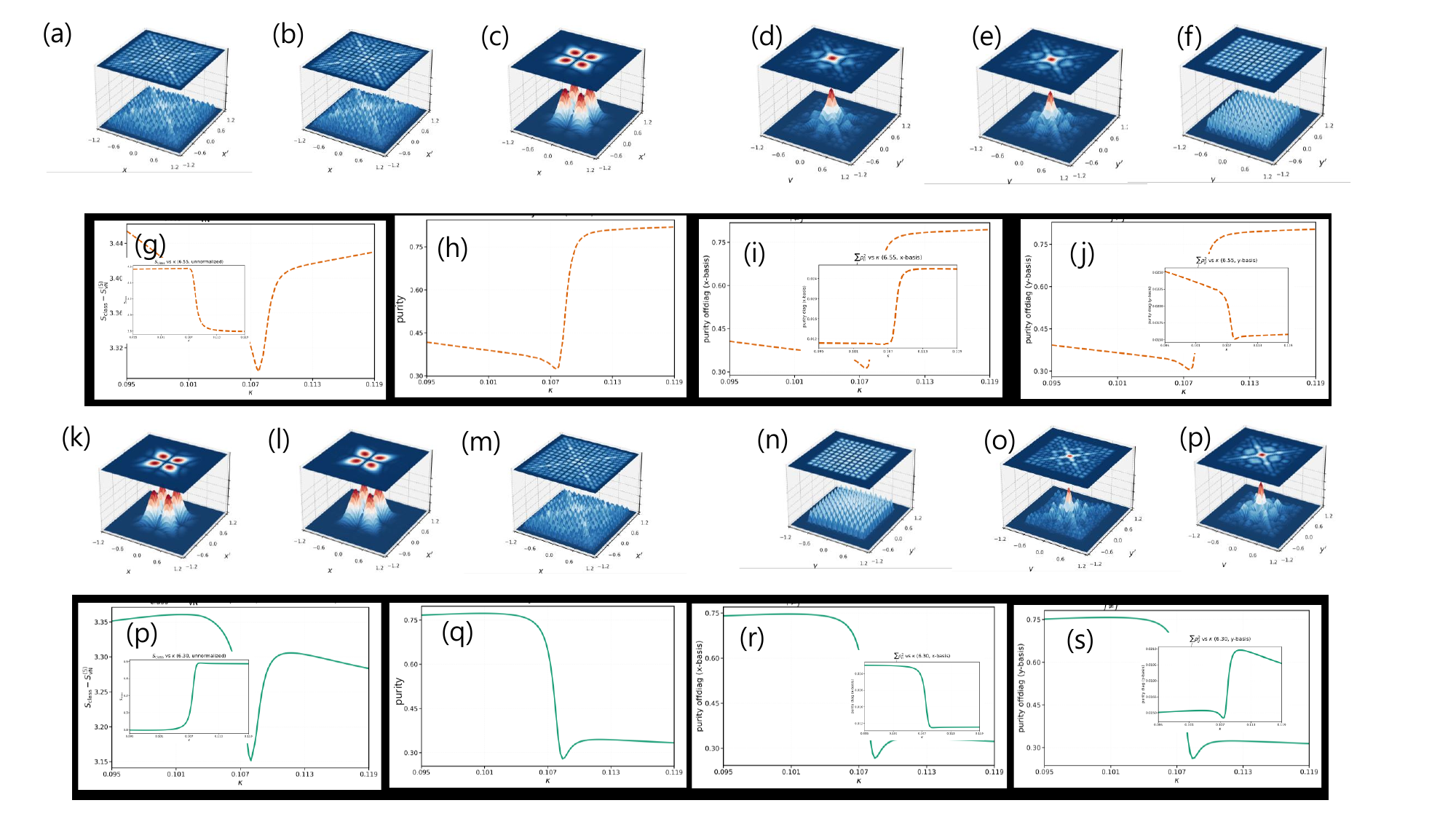}
\caption{\textbf{Quadrupole billiard: reduced-density textures, coherence, and purity across an avoided crossing.}
Reduced matrices retain strong off-diagonal texture, yet \(C_d^{x,y}\) and the purity dip in the hybridization window.
Channel-resolved purity shows that substantial off-diagonal weight persists, indicating an entanglement-driven rebalancing rather than a trivial loss of off-diagonal structure.}
\label{Figure-3}
\end{figure*}
\paragraph{Perturbative sign rule (compact form).}
To connect reshuffling to the entropy trend, we use first-order SVD perturbation theory \cite{StewartSun1990}.
For a small step \(\eta\to\eta+\delta\eta\), write \(\tilde{C}=C+E\).
If \(\sigma_k\) is a simple singular value with left/right singular vectors \(u_k,v_k\), then
\begin{equation}
\Delta\sigma_k \approx \Re\!\left(u_k^\dagger E v_k\right),
\label{eq:delta_sigma_stewart_eta}
\end{equation}
and with \(\lambda_k=\sigma_k^{2}\) the linearized entropy response is weighted by \(1+\ln\lambda_k\), which changes sign at \(\lambda_c=1/e\).
This motivates separating the leading Schmidt weights into an upper sector (\(\lambda_k>\lambda_c\)) and a lower sector (\(\lambda_k<\lambda_c\)), yielding the sign rule summarized in Table~\ref{tab:sign_rule}.

% requires \usepackage{multirow}
\begin{table}[t]
\centering
\caption{Sign rule for the linearized entropy contribution \(\Delta S_k^{(1)}\).}
\label{tab:sign_rule}
\begin{tabular}{c|c|c}
Region & \(r_k=\Re(u_k^\dagger E v_k)\) & \(\mathrm{sgn}(\Delta S_k^{(1)})\)\\
\hline
\multirow{2}{*}{Upper (\(\lambda_k>\lambda_c\))} & \(r_k>0\) & \(-\)\\
 & \(r_k<0\) & \(+\)\\
\hline
\multirow{2}{*}{Lower (\(\lambda_k<\lambda_c\))} & \(r_k>0\) & \(+\)\\
 & \(r_k<0\) & \(-\)\\
\end{tabular}
\end{table}

% =========================================================
% Coherence and purity (Figs. 3 and 4)
% =========================================================
\section{Coherence and purity}
\label{sec:coherence_purity}

The Schmidt-spectrum analysis resolves how hybridization reshapes entanglement.
We now complement that view in a fixed coordinate basis by tracking (i) basis coherence via a resource-theoretic monotone and (ii) a diagonal/off-diagonal decomposition of the Hilbert--Schmidt weight that diagnoses how matrix structure is redistributed.
This combination is essential because the reduced density textures can remain strongly off-diagonal even when the coherence monotone varies sharply; see, e.g., Ref.~\cite{BengtssonZyczkowski2017} for a geometric viewpoint on quantum states and correlations.

\subsection{Relative-entropy coherence from marginal entropies}
\label{sec:relent_coh}

In the resource theory of coherence, coherence is basis dependent and quantified by the entropy increase induced by complete dephasing \cite{Baumgratz2014,StreltsovRMP2017}.
Let \(\Delta[\rho]\) delete all off-diagonal entries in the chosen basis.
The relative entropy of coherence is
\begin{equation}
C_d(\rho)=S(\Delta[\rho]) - S(\rho)\ge 0,
\label{eq:Cd-def}
\end{equation}
with \(S(\rho)=-\mathrm{Tr}(\rho\log\rho)\).

Applying this to \(\rho_x(\eta)\) and \(\rho_y(\eta)\), the diagonal distributions are
\begin{equation}
P_x(i;\eta)=\langle x_i|\rho_x(\eta)|x_i\rangle,\qquad
P_y(j;\eta)=\langle y_j|\rho_y(\eta)|y_j\rangle.
\label{eq:marginals_from_rho_eta}
\end{equation}
Their Shannon entropies are
\begin{equation}
\begin{aligned}
S_{\mathrm{Sh}}^{x}(\eta) &= -\sum_i P_x(i;\eta)\ln P_x(i;\eta),\\
S_{\mathrm{Sh}}^{y}(\eta) &= -\sum_j P_y(j;\eta)\ln P_y(j;\eta).
\end{aligned}
\label{eq:Sh}
\end{equation}
Since \(S(\Delta[\rho_x])\) is exactly the Shannon entropy of \(P_x\) (and similarly for \(y\)),
\begin{equation}
C_d^{x}(\eta)=S_{\mathrm{Sh}}^{x}(\eta)-S(\rho_x(\eta))\;,
C_d^{y}(\eta)=S_{\mathrm{Sh}}^{y}(\eta)-S(\rho_y(\eta)).
\label{eq:Cd_xy_eta}
\end{equation}

For a pure eigenmode, \(S(\rho_x)=S(\rho_y)=S_{\mathrm{vN}}(\eta)\), so \(C_d^{x,y}\) provide basis-resolved complements to entanglement.

\subsection{Purity and Hilbert--Schmidt channel decomposition}
\label{sec:purity_hs}

We also track the purity
\begin{equation}
P(\eta)=\mathrm{Tr}\,\rho_x(\eta)^2=\mathrm{Tr}\,\rho_y(\eta)^2,
\label{eq:purity_def_eta}
\end{equation}
and decompose it in the same reference basis into diagonal and strictly off-diagonal channels,
\begin{equation}
P(\eta)=P_{\mathrm{diag}}(\eta)+P_{\mathrm{off}}(\eta),
\label{eq:purity_decomp_eta}
\end{equation}
with
\begin{equation}
P_{\mathrm{diag}}(\eta)=\sum_i|\rho(i,i)|^2,\qquad
P_{\mathrm{off}}(\eta)=\sum_{i\neq j}|\rho(i,j)|^2.
\label{eq:Pdiag_Poff_eta}
\end{equation}
While \(C_d\) is an entropy-based resource measure, \(P_{\mathrm{diag}}\) and \(P_{\mathrm{off}}\) provide a norm-based diagnostic of how the Hilbert--Schmidt weight is distributed between populations and off-diagonal correlations.

\subsection{Quadrupole versus oval: coherence inversion at avoided crossings}
\label{sec:coh_results}

\begin{figure*}
\centering
\includegraphics[width=16.5cm]{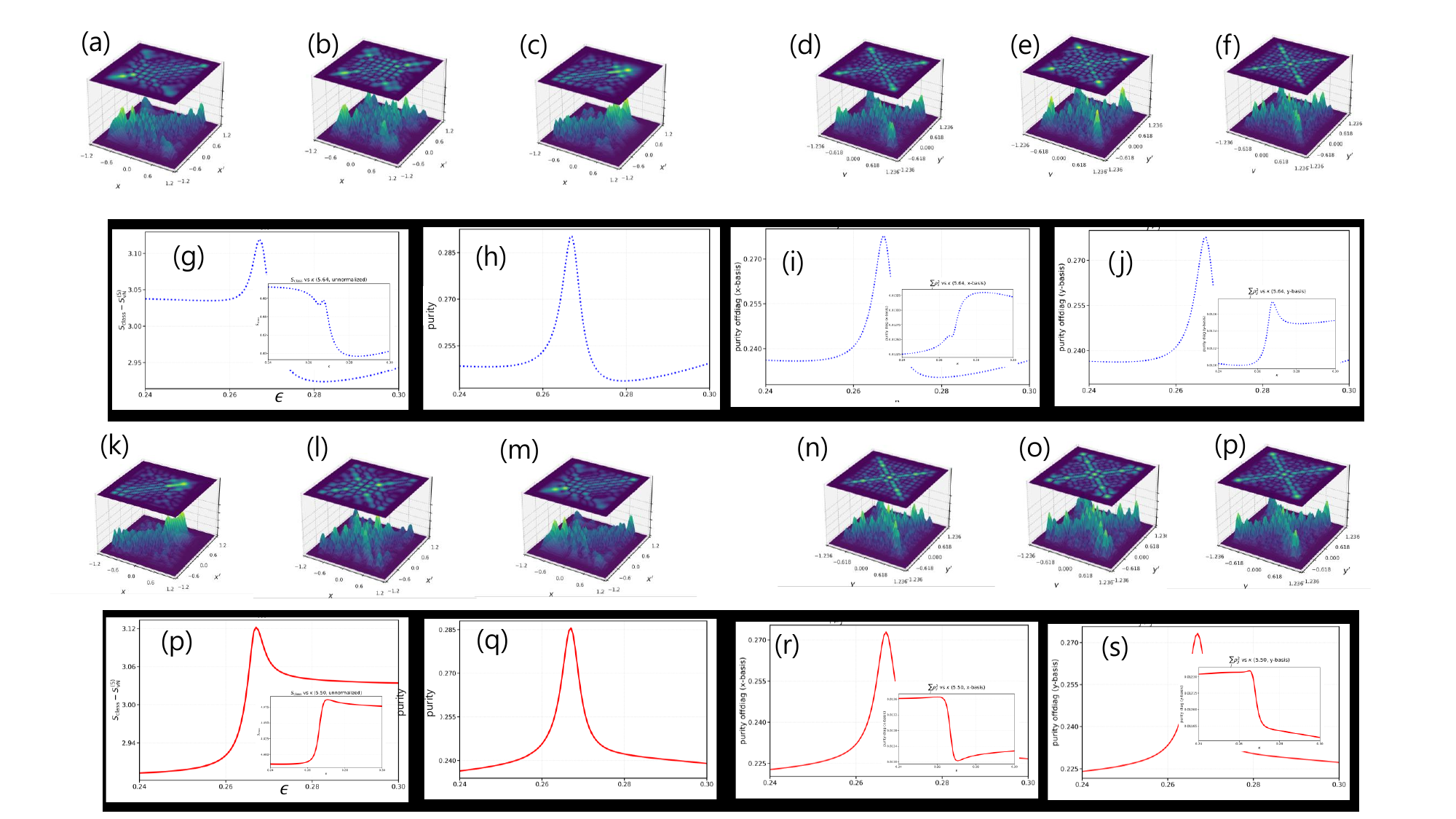}
\caption{\textbf{Oval billiard: the same diagnostics as Fig.~\ref{Figure-3} with an inverted response.}
Despite similarly strong off-diagonal texture, \(C_d^{x,y}\) and the purity peak in the hybridization window, consistent with the entanglement dip in Fig.~\ref{Figure-2}(f).
Off-diagonal dominance persists, so the coherence peak reflects a genuine reorganization of correlations rather than a transition to a purely diagonal reduced state.}
\label{Figure-4}
\end{figure*}

Figures~\ref{Figure-3} and \ref{Figure-4} resolve the basis-dependent structure behind Fig.~\ref{Figure-2}.
In the quadrupole billiard, \(C_d^{x,y}\) exhibits a clear dip accompanied by a purity dip, indicating that the reduced state becomes more mixed at hybridization (entanglement enhancement) and that the entropy increase of \(\rho_{x,y}\) outpaces the increase of the diagonal (marginal) entropies.
In the oval billiard, the response is inverted: \(C_d^{x,y}\) peaks together with a purity peak, mirroring the entanglement dip in Fig.~\ref{Figure-2}.
Crucially, the channel-resolved purity confirms in both geometries that off-diagonal weight remains substantial, so the coherence inversion reflects an entanglement-driven redistribution of correlations rather than the disappearance of off-diagonal structure.

Taken together with Fig.~\ref{Figure-1}, these results show that configuration-space delocalization and Landau--Zener-type mode exchange do not uniquely fix the information-theoretic response: entanglement and coherence provide an independent axis that can distinguish chaos realizations that are otherwise similar by conventional wave-chaos diagnostics.

% =========================================================
% Ising model (Figs. 5--7)
% =========================================================
\section{Quantum-informational analysis in the tilted-field Ising chain}
\label{sec:model-protocol}

The transverse-field Ising chain is integrable in the purely transverse-field limit \cite{Pfeuty1970}; adding a longitudinal component via a field tilt breaks integrability and can drive the onset of quantum chaos in spin chains \cite{SantosRigol2010}. In such nonintegrable many-body systems, chaotic eigenstate structure is closely tied to thermalization and eigenstate thermalization \cite{Rigol2008,DAlessio2016}.

\subsection{Model, symmetry sector, and branch tracking}
We consider an open spin-\(\tfrac12\) Ising chain in a tilted field,
\begin{equation}
H(\theta)=J\sum_{n=1}^{L-1}\sigma_n^{z}\sigma_{n+1}^{z}
+B\sum_{n=1}^{L}\bigl(\sin\theta\,\sigma_n^{x}+\cos\theta\,\sigma_n^{z}\bigr),
\label{eq:tfim_tilt}
\end{equation}
with fixed parameters
\begin{equation}
L=8,\qquad J=1,\qquad B=1,
\label{eq:params_fixed}
\end{equation}
so that the global tilt \(\theta\) is the sole control parameter.
To avoid mixing symmetry sectors we work in the bit-reversal even subspace (dimension \(136\) for \(L=8\)) and track adiabatic branches by overlap maximization between neighboring \(\theta\) steps.

\begin{figure*}
\centering
\includegraphics[width=16.5cm]{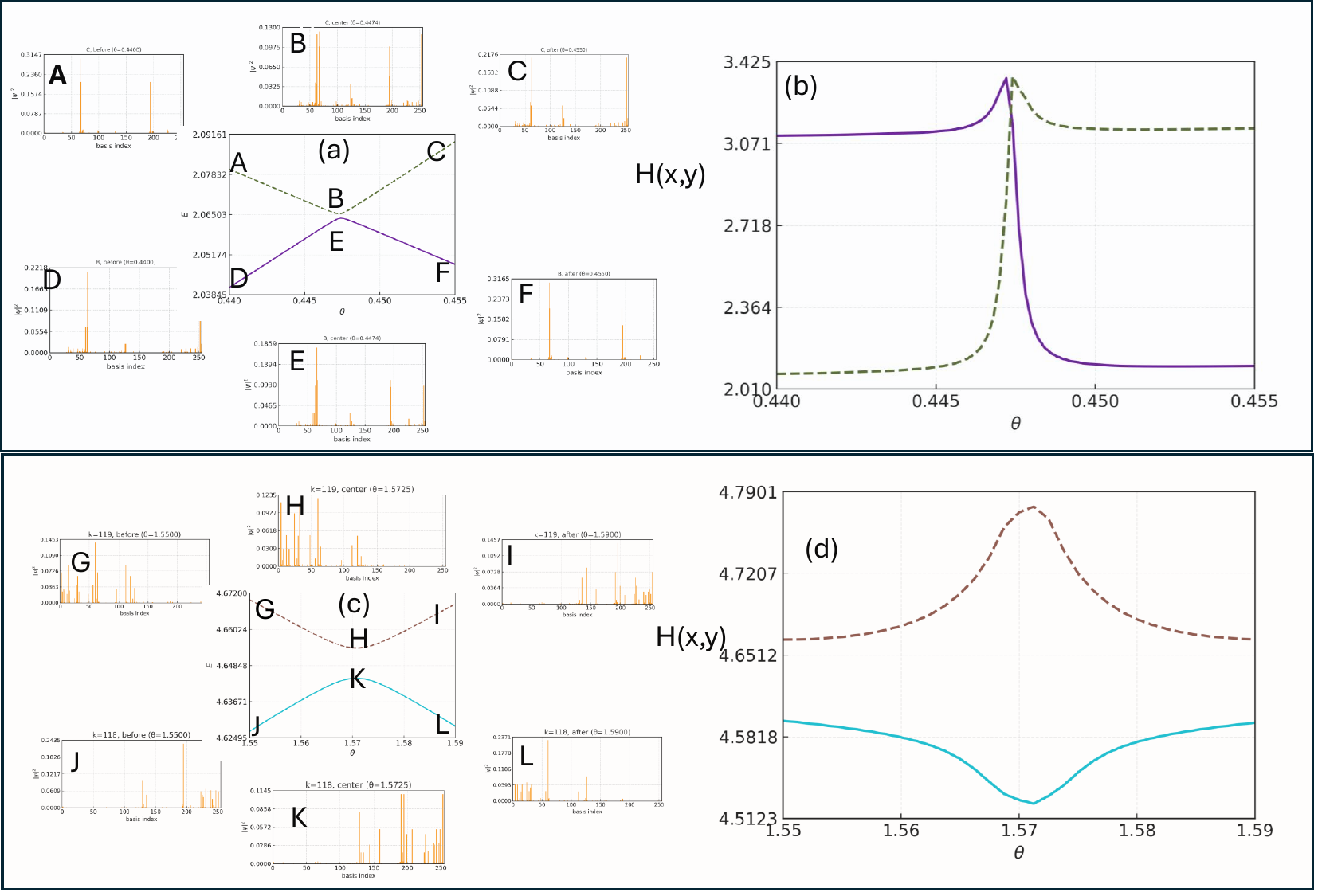}
\caption{\textbf{Ising chain: avoided crossings and computational-basis delocalization.}
(a,b) Soft-chaos window near \(\theta\simeq 0.45\): an avoided crossing produces rapid basis reshuffling and a localized peak in the configuration-space Shannon entropy.
(c,d) Strong-chaos window near \(\theta\simeq 1.57\): baseline delocalization is already high and the avoided-crossing response becomes branch dependent (peak versus dip).}
\label{Figure-5}
\end{figure*}

\subsection{Configuration-space delocalization}
Expanding \(|\psi(\theta)\rangle=\sum_{n}\psi_n(\theta)|n\rangle\) in the computational \(\sigma^z\) basis, we define
\begin{equation}
S_{\mathrm{config}}(\theta)=-\sum_{n}|\psi_n(\theta)|^2\log|\psi_n(\theta)|^2.
\label{eq:Sconfig}
\end{equation}
Figure~\ref{Figure-5} shows that the avoided crossing enhances \(S_{\mathrm{config}}\) in the soft-chaos window, while in the strong-chaos window the response is mode dependent, consistent with hybridization on top of an already strongly mixed background.

\subsection{Bipartition, reduced states, and entanglement}
We split the chain into two halves \(A\) (sites \(1\ldots 4\)) and \(B\) (sites \(5\ldots 8\)),
\begin{equation}
\mathcal{H}=\mathcal{H}_A\otimes\mathcal{H}_B,\qquad
\mathcal{H}_A\simeq(\mathbb{C}^2)^{\otimes 4},\quad
\mathcal{H}_B\simeq(\mathbb{C}^2)^{\otimes 4}.
\label{eq:HA_HB_split}
\end{equation}
Writing
\begin{equation}
|\psi(\theta)\rangle=\sum_{a=0}^{15}\sum_{b=0}^{15}\psi_{ab}(\theta)\,|a\rangle_A\otimes|b\rangle_B,
\label{eq:psi_AB_expansion}
\end{equation}
the reduced states are
\begin{equation}
\rho_A(\theta)=\mathrm{Tr}_B\!\left(|\psi(\theta)\rangle\langle\psi(\theta)|\right),\;
\rho_B(\theta)=\mathrm{Tr}_A\!\left(|\psi(\theta)\rangle\langle\psi(\theta)|\right),
\label{eq:rhoA_rhoB_def}
\end{equation}
with matrix elements
\begin{equation}
[\rho_A]_{aa'}=\sum_{b}\psi_{ab}\psi_{a'b}^*,\qquad
[\rho_B]_{bb'}=\sum_{a}\psi_{ab}^*\psi_{ab'}.
\label{eq:rhoA_elements}
\end{equation}
The Schmidt coefficients are the nonzero eigenvalues of \(\rho_A\) and the entanglement entropy is
\begin{equation}
S_{\mathrm{vN}}(\theta)=-\sum_i \lambda_i(\theta)\ln \lambda_i(\theta).
\label{eq:SvN_ising}
\end{equation}

\begin{figure*}
\centering
\includegraphics[width=17.5cm]{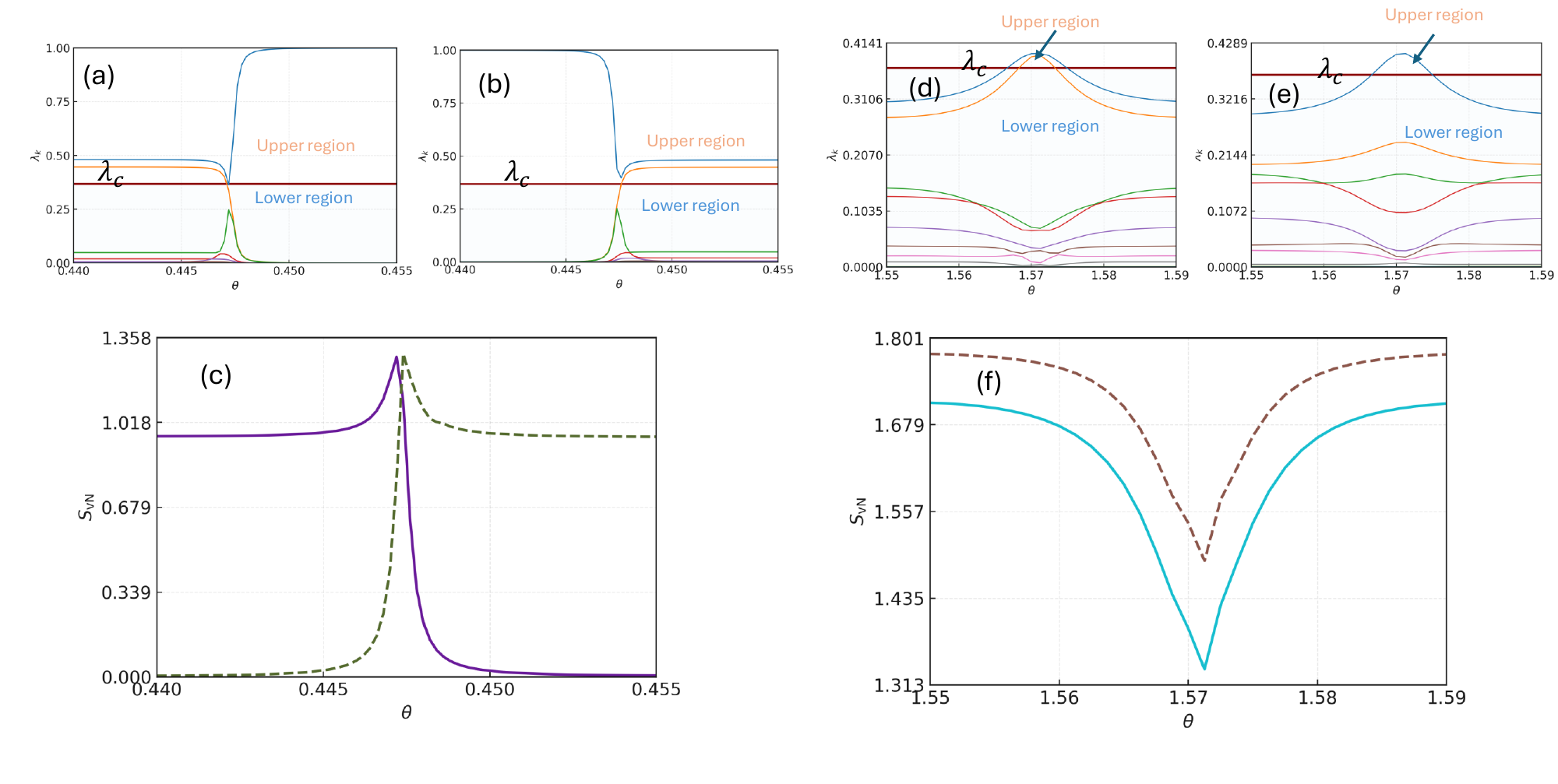}
\caption{\textbf{Ising-chain analogue of Fig.~\ref{Figure-2}: Schmidt-spectrum reshuffling and entanglement inversion.}
(a--c) Soft-chaos window: hybridization redistributes Schmidt weight into subdominant components and produces an entanglement peak with a clear branch-exchange pattern.
(d--f) Strong-chaos window: hybridization locally concentrates the spectrum and produces an entanglement dip on top of a large baseline.}
\label{Figure-6}
\end{figure*}

Figure~\ref{Figure-6} reproduces the billiard dichotomy within a single Hamiltonian.
In the soft-chaos window, the avoided crossing produces a sharp entanglement peak with a pronounced exchange between partner branches.
In the strong-chaos window, the avoided crossing yields an entanglement dip instead; notably, the dip occurs on top of an already large background entanglement, and even at the minimum \(S_{\mathrm{vN}}\) remains of order unity (Fig.~\ref{Figure-6}f), so coherence-sensitive structure can be reshaped without quenching bipartite correlations.

\subsection{Reduced-density textures and purity channels}
We track mixedness via the purity
\begin{equation}
P(\theta)=\mathrm{Tr}\,\rho_A(\theta)^2=\mathrm{Tr}\,\rho_B(\theta)^2,
\label{eq:purity_ising}
\end{equation}
and decompose it into diagonal and off-diagonal Hilbert--Schmidt channels in the computational basis of the half-chain,
\begin{equation}
P(\theta)=P_{\mathrm{diag}}(\theta)+P_{\mathrm{off}}(\theta),
\label{eq:purity_decomp_ising}
\end{equation}
with \(P_{\mathrm{diag}}=\sum_i|\rho_{ii}|^2\) and \(P_{\mathrm{off}}=\sum_{i\neq j}|\rho_{ij}|^2\).

\begin{figure*}
\centering
\includegraphics[width=16.5cm]{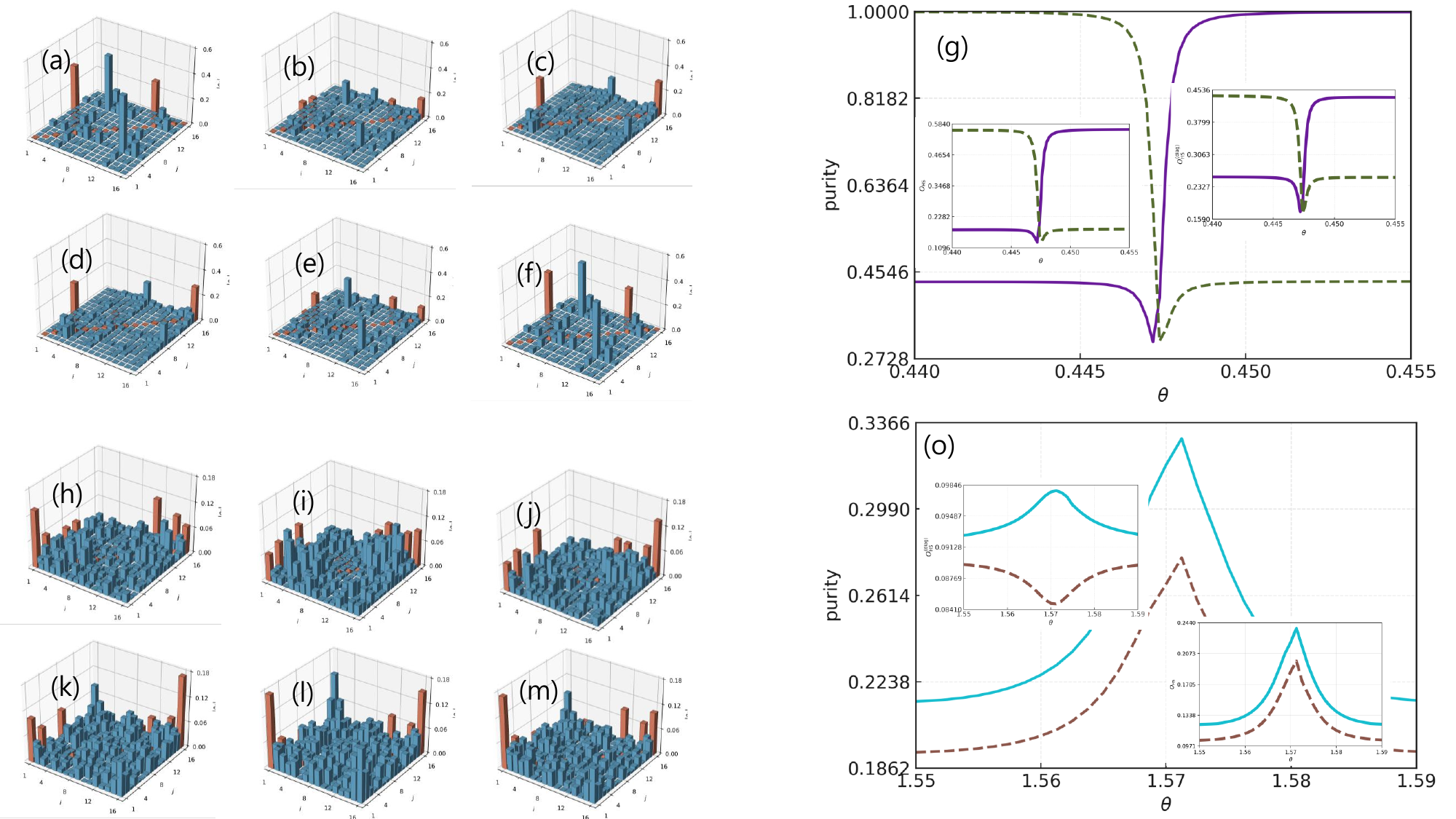}
\caption{\textbf{Ising chain: reduced-density structure and purity-channel competition across avoided crossings.}
(a--g) Soft-chaos window: \(|(\rho_A)_{ij}|\) is relatively structured and the purity dips at the avoided crossing; diagonal and off-diagonal channels contribute comparably.
(h--o) Strong-chaos window: \(|(\rho_A)_{ij}|\) is broadly populated and the purity peaks at the avoided crossing; for Mode~1, \(P_{\mathrm{off}}\) peaks while \(P_{\mathrm{diag}}\) forms a dip, demonstrating channel competition in a strongly mixed background.}
\label{Figure-7}
\end{figure*}

Figure~\ref{Figure-7} highlights a key many-body refinement beyond the billiards.
In the strong-chaos window, the avoided crossing can increase off-diagonal correlations even when the diagonal sector simultaneously dips, so the same total purity feature can arise from qualitatively different internal reallocations between population and coherence-like channels.

% =========================================================
% Operational outlook (compressed, no "minimal check")
% =========================================================
\subsection{Operational outlook: one-knob control of antagonistic resources}
\label{sec:ising_operational_outlook}

From a resource-theoretic viewpoint, coherence is basis dependent and the relative-entropy coherence is a coherence monotone with operational meaning as the asymptotic distillable coherence rate \cite{Baumgratz2014,StreltsovRMP2017,WinterYang2016}.
Moreover, coherence is operationally linked to entanglement through convertibility under incoherent operations assisted by an incoherent ancilla \cite{Streltsov2015}, and coherence monotones admit direct interpretations in discrimination settings \cite{Napoli2016}.

With this in mind, Figs.~\ref{Figure-6} and \ref{Figure-7} show that tuning a single microscopic parameter, the global tilt \(\theta\), toggles between two qualitatively distinct resource responses of hybridization in the same Hamiltonian: an entanglement-enhancing, coherence-suppressing response in a soft-chaos window and an inverted response in a strong-chaos window.
The strong-chaos regime is particularly notable because the entanglement dip occurs on top of a large baseline entanglement, while the reduced state exhibits a sharp and structured reallocation between diagonal and off-diagonal purity channels (channel competition).
Operationally, this resonates with a broader lesson from quantum technologies: useful quantum advantage can persist in strongly mixed or noisy settings when off-diagonal structure is selectively preserved or rebalanced, as exemplified by coherence-enabled interference protocols \cite{Scully1989,Fleischhauer2005} and mixed-state quantum information primitives \cite{KnillLaflamme1998,Datta2008,Tan2008,Rebentrost2009}.
Our results therefore position chaotic hybridization as a practical mechanism for task-oriented quantum resource shaping, with direct experimental relevance in wave-chaos resonators and programmable spin simulators \cite{Joseph2015,Joseph2016,Islam2015,Elben2018}.

% =========================================================
% Conclusion
% =========================================================
\section{Discussion and conclusion}

We have shown that chaotic dynamics can serve as a \emph{resource switch}: localized hybridization windows around A.C.s selectively enhance either entanglement or basis coherence within the same platform. Physically, an A.C. acts as a local hybridization valve that reshuffles Schmidt weights and redistributes coherence between diagonal and off-diagonal sectors of reduced states. A central implication is a stronger universality: the \emph{direction} of resource flow across A.C.s can be robust across inequivalent subsystem notions, spanning inter-degree-of-freedom entanglement in single-wave settings and genuine inter-particle entanglement in many-body settings.

In two wave-chaotic billiards, A.C.s produce comparable mode exchange and configuration-space delocalization peaks, yet the entanglement response inverts between geometries: one shows Schmidt-spectrum equalization and an entanglement peak, while the other shows spectral concentration and an entanglement dip. This inversion persists when extended to basis-resolved coherence and purity-channel measures, confirming that delocalization alone does not fix the information-theoretic structure of eigenmodes. The tilted-field Ising chain provides the many-body counterpart within one Hamiltonian: the global tilt angle selects soft- versus strong-chaos windows and toggles between entanglement-enhancing and coherence-enhancing A.C. responses, with a diagonal/off-diagonal channel competition in the reduced state.

Experimentally, the billiard diagnostics require only spatial eigenmode data on a grid to construct reduced states and Schmidt spectra \cite{Joseph2015,Joseph2016,Stockmann1999,NockelStone1997}, while the many-body quantities are accessible via subsystem measurements in programmable spin platforms \cite{Islam2015,Elben2018}. More broadly, single-knob switching between entanglement-oriented and coherence-oriented regimes suggests a route to task-dependent state selection in quantum simulators and wave-based devices.

\section{acknowledgement}
This work was supported by the National Research Foundation of Korea (NRF) through a grant funded by the Ministry of Science and ICT (Grants Nos. RS2023-00211817 and RS-2025-00515537), the Institute for Information \& Communications Technology Promotion (IITP) grant funded by the Korean government (MSIP) (Grants Nos. RS-2019-II190003 and RS-2025-02304540), the National Research Council of Science \& Technology (NST) (Grant No. GTL25011-000), and the Korea Institute of Science and Technology Information (KISTI). S.L. acknowledges support from the National Research Foundation of Korea (NRF) grants funded by the MSIT (Grant No. RS-2022-NR068791).

% =========================================================
% Bibliography (bibitem style)
% =========================================================

\end{document}